\documentclass[]{aa}
\usepackage{times}
\usepackage{natbib}
\usepackage[dvips]{graphicx}
\usepackage{psfig}
\bibpunct{(}{)}{;}{a}{}{,}

\def\pdot {\dot P}

\def\ltsima{$\; \buildrel < \over \sim \;$}
\def\lsim{\lower.5ex\hbox{\ltsima}}
\def\gtsima{$\; \buildrel > \over \sim \;$}
\def\gsim{\lower.5ex\hbox{\gtsima}}
\def\msun{~M_{\odot}}

\def\XMM{{\em XMM--Newton}}
\def\RXTE{{\em RXTE}}
\def\SAX{{\em BeppoSAX}}
\def\saxj{SAX J0635.2+0533}

\def\Swift{{\em Swift}}

\begin{document}

\title
{\Swift\ monitoring of the massive X-ray binary \\ \saxj}

\author{Nicola La Palombara\inst{1}, Sandro Mereghetti\inst{1}}

\institute {INAF, Istituto di Astrofisica Spaziale e Fisica
Cosmica Milano, via E.\ Bassini 15, I-20133 Milano, Italy }

\offprints{N. La Palombara, nicola@iasf-milano.inaf.it}

\date{Received / Accepted}

\authorrunning{N. La Palombara \& S. Mereghetti}

\titlerunning{X-ray pulsar \saxj}

\abstract{\saxj\ is a binary pulsar with a very short pulsation period ($P$ = 33.8 ms) and a high long-term spin down ($\pdot$ $>$ 3.8$\times10^{-13}$ s s$^{-1}$), which suggests a rotation-powered (instead of an accretion-powered) nature for this source. While it was discovered at a flux level around 10$^{-11}$ erg cm$^{-2}$ s$^{-1}$, between 2003 and 2004 this source was detected with \XMM\ at an average flux of about 10$^{-13}$ erg cm$^{-2}$ s$^{-1}$; moreover, the flux varied of over one order of magnitude on time scales of a few days, sometimes decreasing below $3\times10^{-14}$ erg cm$^{-2}$ s$^{-1}$. Since both the rotation-powered and the accretion-powered scenarios have difficulties to explain these properties, the nature of \saxj\ is still unclear.
Here we report on our recent long term monitoring campaign on \saxj\ carried out with \Swift\ and on a systematic reanalysis of all the \RXTE\ observations performed between 1999 and 2001. We found that during this time interval the source remained almost always active at a flux level above 10$^{-12}$ erg cm$^{-2}$ s$^{-1}$. 
%

}

\maketitle

\section{Introduction}

\saxj\ (also known as PSR J0635+0533) is a very peculiar X-ray source. Its first detection, obtained with \SAX\ in 1997 \citep{kaa99}, revealed a 2-10 keV flux of 1.2 $\times$ 10$^{-11}$ erg cm$^{-2}$ s$^{-1}$ and a hard power-law spectrum (photon index $\sim$1.5) extending to 40 keV; based on these spectral properties and on the positional coincidence with a Be star, it was classified as a high-mass X-ray binary (HMXRB). The \SAX\ data revealed also X--ray pulsations at 33.8 ms \citep{cus00}. This is the shortest pulsation period observed in a high-mass binary pulsar: a comparably short period in a HMXRB has been detected only in A0538-66, a transient Be/X-ray binary in the Large Magellanic Cloud with a pulse period of 69 ms \citep{ski82}. Based on the pulse frequencies measured with \RXTE\ in 1999, it was possible to set a lower limit ($\pdot$ $>$ 3.8$\times10^{-13}$ s s$^{-1}$) on its long--term spin down \citep{kaa00}. This slow-down rate is high enough that the corresponding rotational energy loss (${\dot E}_{\rm rot}$ = 10$^{45}$ 4$\pi^{2}$ $\pdot /P^3$ = 4$\times$10$^{38}$ erg s$^{-1}$) can sustain the observed X-ray luminosity: in this interpretation, alternative to the mass-accretion scenario, the X-ray emission could be due to the shocks between the relativistic wind of the pulsar and that of the companion star. This emission mechanism has been invoked for other binary systems composed of a compact object (a neutron star in most cases) orbiting a massive star, such as PSR B1259--63 \citep{Tam+15}, LS I +61$^\circ$ 303 \citep{Paredes-Fortuny+15}, and LS 5039 \citep{Takata+14}. These binaries have been observed also at very high energies, up to the TeV range, thus proving that they are among the main sites of particle acceleration in the Galaxy \citep{Dubus15}. It is interesting to note that \saxj\ was considered a candidate counterpart of the \textit{EGRET} source 2EG J0635+0521/3EG J0634+0521 \citep{Hartman+99}, since its position is within the error box of this unidentified galactic $\gamma$--ray source. This \textit{EGRET} source corresponds to the \textit{FERMI} source 1FGL J0636.0+0458c \citep{Abdo+10}, which however was not confirmed in the second \citep{Nolan+12} and third \citep{Acero+15} \textit{FERMI} catalogues; very probably in these catalogues the source was undetected since its significance dropped below the threshold, either as a result of time variability, change in the diffuse model, or the shift from unbinned to binned likelihood in the catalog analysis procedure.

Between September 2003 and April 2004 \saxj\ was observed by \XMM\ in two different periods, with 8 pointings in September--October 2003 (epoch D) and other 2 pointings in March--April 2004 (epoch E, \citet{MereghettiLaPalombara09}). The source was detected only in six pointings, showing a large variability among the different observations: between September and October 2003 the source flux varied by at least a factor 10, with a rise/decay time of only a few days. The source spectrum was well fitted with an absorbed power law and the average flux in the energy range 0.2--12 keV was $f_{\rm X}\sim 1.5\times10^{-13}$ erg cm$^{-2}$ s$^{-1}$. The maximum flux detected in these observations was $\sim 5 \times$10$^{-13}$ erg cm$^{-2}$ s$^{-1}$, a factor $>$ 20 smaller than that measured at the time of the discovery in 1997. For the other four observations we could only set an upper limit on the source flux, with a minimum upper-limit value $f_{\rm X} < 3\times10^{-14}$ erg cm$^{-2}$ s$^{-1}$.

There are no precise estimates for the distance of \saxj, but from the properties of the proposed optical counterpart \citet{kaa99} derived a range of 2.5--5 kpc; a distance far in excess 5 kpc is unlikely, since this source is in the Galactic anti--center direction. Assuming a conservative source distance d$_5$ of 5 kpc, the average flux during the \XMM\ observations corresponds to a luminosity of a few 10$^{32}$ d$_{5}^{2}$ erg s$^{-1}$, a value which is very small compared to the classical Be/neutron stars systems.

Due to the low luminosity and short spin period, direct accretion of matter onto the neutron star surface would require a magnetic field smaller than B $\sim 10^8$ G, at least three orders of magnitude lower than that expected in a young neutron star. If instead \saxj\ has a typical magnetic field, its low luminosity observed with \XMM\ could be explained with mass accretion stopped at the magnetospheric radius \citep{Campana+95}. However, this scenario is difficult to reconcile with the higher luminosity state observed with \SAX\ and \RXTE, since the pulsations with a relatively high pulsed fraction are unlikely when the magnetic centrifugal barrier operates.

Also the alternative scenario of a rotation--powered neutron star presents some problems. The large flux variability seen in September 2003 could be explained only if the source was near to periastron in a very eccentric orbit, where big changes in the shock properties can be expected. Moreover, if the system indeed contains a neutron star with ${\dot E}_{\rm rot}$ greater than a few 10$^{38}$ erg s$^{-1}$, a larger pulsar luminosity would be expected, since the typical efficiency of conversion of rotational energy to X-ray luminosity is of the order of 10$^{-3}$, as observed in the case of PSR B1259--63 \citep{cer06,cer09} and PSR B1957+20 \citep{hua07}.

In summary, \saxj\ is a unique source among the HMXRBs. Its luminosity is unusually low, both in quiescence and during outbursts. Due to these properties, both the accretion--powered and the rotation--powered scenarios imply several peculiarities with respect to other known HMXRBs. In order to investigate the long-- and short--term behaviour of this source, we performed a monitoring campaign with \Swift\ between 2015 and 2016 and analyzed in a systematic way all the observations performed with \RXTE\ between 1999 and 2001. In this way, we could obtain the overall long-term light-curve of this source and study its variability pattern. Our aim was to investigate if this source is characterized by prolonged periods of activity or it spends most of the time in a quiescent state, interrupted only sporadically by bright (possibly periodic) outbursts.

\section{Observations and data analysis}
\label{obs}
\subsection{\Swift}
\label{obs_swift}

\saxj\ was observed repeatedly with \Swift: 31 different pointings were performed between 2015 November 12$^{th}$ and 2016 May 15$^{th}$ (epoch F); then, after a time gap of three months, two additional pointings were performed on 2016 August 16$^{th}$ and 23$^{rd}$ (epoch G). We considered the data of the X-ray Telescope (XRT), which were obtained in photon-counting mode \citep{Gehrels+04}. In Table~\ref{observations_swift} we report the date and effective exposure time of each observation.

We ignored pointing 28, since it was too short ($\sim$ 100 s) for a meaningful investigation. For each of the other pointings we retrieved from the \Swift\ archive\footnote{http://www.swift.ac.uk/swift\_live/index.php} the reduced files produced with the standard \textsc{xrtpipeline} processing routine. To study the source variability, we analyzed separately the 32 data-sets; for each of them we performed a source detection to obtain an accurate evaluation of the source count rate (CR) and significance. We considered the source detected if its \textit{signal-to-noise} ratio was above a threshold value of 2 $\sigma$ c.l.; we adopted such a loose detection criterion considering that the presence of a source at this position was already known. We found that \saxj\ was detected in 22 observations, while it was undetected in the first six observations (December 2015) and in other four observations during 2016; in case of a non-detection we set a 3 $\sigma$ c.l. upper limit on the CR (Table~\ref{observations_swift}).

\begin{figure}[h]
\includegraphics[angle=0,height=8cm]{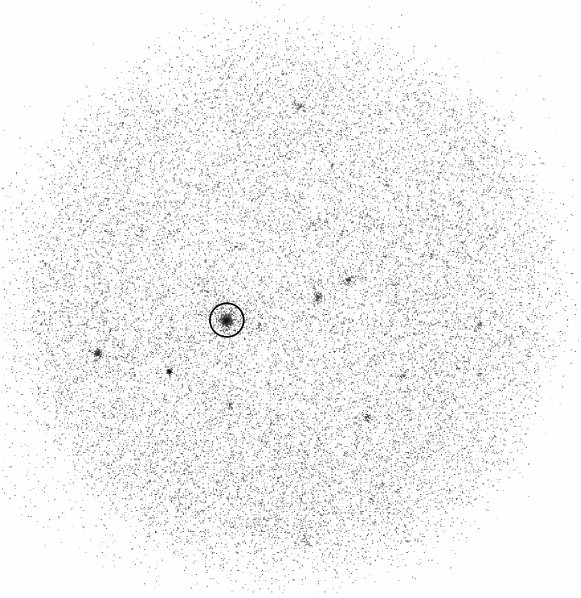}
\caption{Image of the sky region around the position of \saxj\ (black circle), obtained by summing the 22 \Swift/XRT observation where the source was detected; the sky region has a radius of 13 arcmin, while the black circle has a radius of 47 arcsec.}\label{swift_image}
\end{figure}

\begin{figure}[h]
\centering
\resizebox{\hsize}{!}{\includegraphics[angle=-90,clip=true]{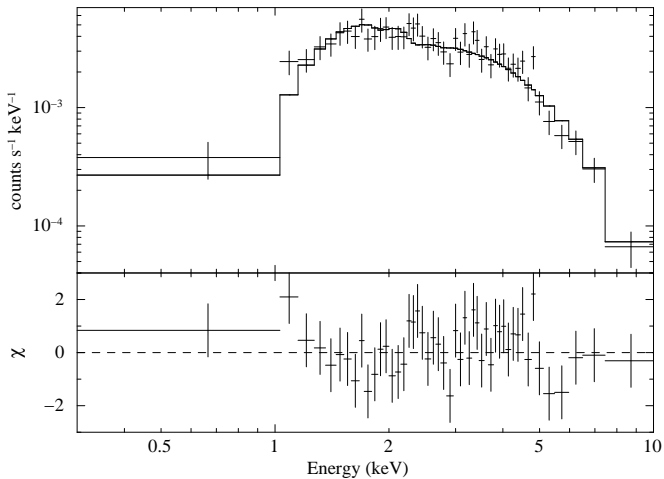}}
\caption{\textit{Top panel}: average spectrum of \saxj\ with the best--fit power--law model. \textit{Bottom panel}: data-model residuals, in units of $\sigma$.}\label{spectra}
\end{figure}

In order to investigate the average spectral properties of the source, we used \textsc{xselect} to merge the event lists of the 22 observations where the source was detected and to accumulate an image of the field--of--view (reported in Fig.\ref{swift_image}). In the total observing time of $\sim$ 74 ks, \saxj\ was detected with an average CR = (1.58$\pm$0.05)$\times 10^{-2}$ cts s$^{-1}$. To minimize the background contribution, we extracted the source spectrum by selecting events in a circular region of 20 pixel radius, corresponding to $\sim$ 47$''$; the background spectrum was accumulated from a large circular area with no sources and radius of 100 pixels. We generated a cumulative exposure image of the merged observations and, then, used the task \textsc{xrtmkarf} to calculate the ancillary response file. The total spectrum was rebinned with a minimum of 20 counts per bin, and fitted in the energy range 0.3--10 keV using \textsc{XSPEC} 12.7.0. We obtained a good fit with an absorbed power--law (Fig.\ref{spectra}), yielding a hydrogen column density $N_{\rm H} = (2.7^{+0.6}_{-0.5})\times10^{22}$ cm$^{-2}$ and a photon index $\Gamma=2.0^{+0.2}_{-0.3}$. The average absorbed flux in the energy range 0.3--10 keV is $f_{\rm X} = (1.08^{+0.04}_{-0.08})\times10^{-12}$ erg cm$^{-2}$ s$^{-1}$, while the corresponding unabsorbed flux is $2.5\times10^{-12}$ erg cm$^{-2}$ s$^{-1}$. Acceptable fits were also obtained with a thermal bremsstrahlung (\textit{kT} = 6.5 keV) and with a blackbody (\textit{kT} $\simeq$ 1 keV), although in the last case there are large residuals at high energies.

\begin{figure}[htp!]
\begin{center}
\resizebox{\hsize}{!}{\includegraphics[angle=-90,clip=true]{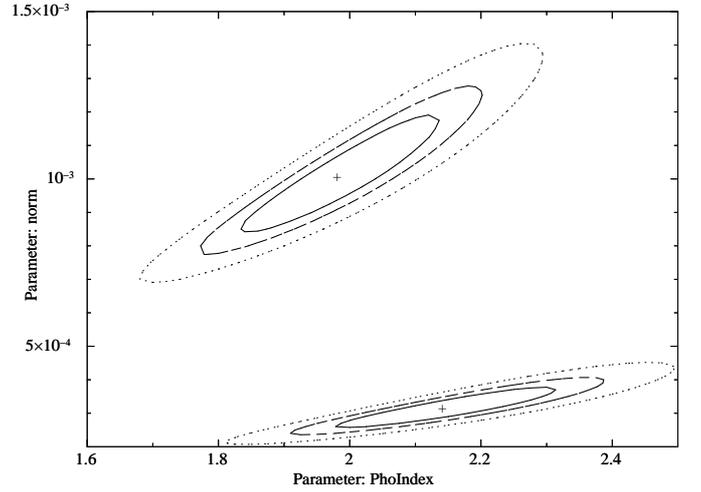}}
\end{center}
\vspace{-0.5cm}
\caption{Contour plots between the power-law photon index and normalization in the case of the highest and the lowest flux levels for \saxj.}
\label{contour_plot}
\end{figure}

To investigate also the possible spectral dependence on the source flux, we accumulated two separate spectra for, respectively, the observations with a source CR $>$ 0.02 cts s$^{-1}$ and those with a lower CR; in this way we obtained two spectra with a similar count statistics. We performed a simultaneous fit of the two spectra with an absorbed power-law model, assuming a common hydrogen column density in order to avoid any degeneracy with the photon index. In Fig.\ref{contour_plot} we report, for both spectra, the contour plot between the normalization and the photon index, which shows no evidence for a variation in the photon index within the uncertainties. Using the flux/CR ratio obtained for the total spectrum we converted the measured CR of each observation (or, in the case of missing detection, its upper limit at a 3 $\sigma$ c.l.) into the corresponding energy flux; they are reported in Fig.~\ref{all}.

\begin{figure}[htp!]
\begin{center}
\resizebox{\hsize}{!}{\includegraphics[angle=-90,clip=true]{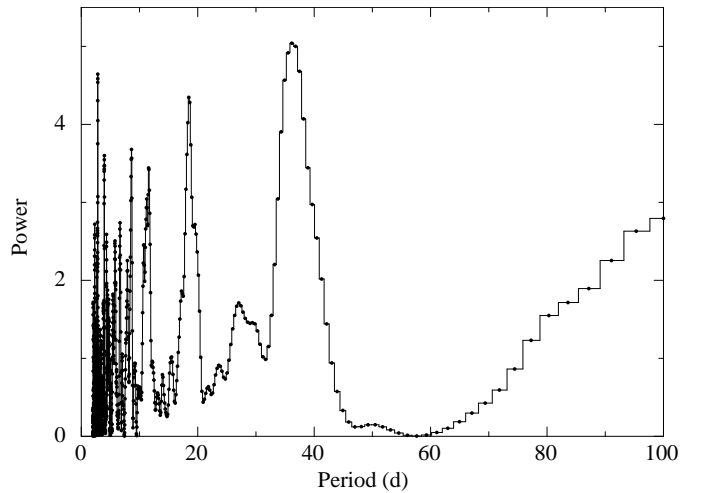}}
\end{center}
\vspace{-0.5cm}
\caption{Periodogram of \saxj\ obtained through the Lomb--Scargle analysis of the \Swift\ data obtained between 2015 and 2016.}
\label{periodogram}
\end{figure}

We also performed a search for long-term periodicities in the \Swift\ data by applying the Lomb--Scargle method \citep{Scargle82}. In Fig.~\ref{periodogram} we report, as an example, the power distribution obtained by searching 1000 independent periods. In this case, a power $>$ 11.5 is necessary to claim a signal detection at 99 \% c.l.; on the other hand, the highest peak in our diagram, with power $\simeq$ 5, has a probability higher than 99.9 \% to be obtained in the absence of a periodic signal. Therefore, we found no evidence of a periodic signal; in particular, with the available data it is not possible to confirm the proposed orbital period of 11.2 days \citep{kaa00}.

\subsection{\RXTE}
\label{obs_rxte}

\RXTE\ \citep{Bradt+93} observed \saxj\ in three different epochs between 1999 and 2001, with 13 pontings in August--September 1999 (epoch A), 5 pointings in May 2000 (epoch B), and 49 pointings between November 2000 and January 2001 (epoch C); in Table~\ref{observations_rxte} we report the dates and effective exposure times of each observation. Only the results of epoch A were published \citep{kaa00}. Therefore, we have analyzed in the same way all the observations of the three epochs. We considered only the data of the Proportional Counter Array (PCA), which covers the energy range 2--100 keV and comprises five identical coalligned gas-filled Proportional Counter Units (PCUs). The exposures of the individual observations were between 1.6 and 6.9 ks and, during each of them, between 2 and 5 PCUs were active.

For the spectral analysis we considered the Standard 2 data and applied the procedure suggested by the instrument team\footnote{http://heasarc.gsfc.nasa.gov/docs/xte/recipes/pca\_spectra.html}. We analyzed separately each observation; moreover, within each observation, we analyzed separately the time intervals characterized by a different number of active PCUs (if any). In each case we created the applicable Good Time Interval and extracted the source and background spectra using only the PCU top layer, which gives a higher signal-to-noise ratio. Then we corrected for the dead time and created the applicable response matrix.

Although the source was always detected, for most observations the spectrum had a low count statistics, which hampered an accurate spectral analysis. Therefore, we assumed an absorbed power-law model fixing both $N_{\rm H}$ and $\Gamma$ to the best-fit values obtained with \Swift\ (2.7$\times10^{22}$ cm$^{-2}$ and 2, respectively) and left only the normalization free to vary. In this way we estimated the fluxes reported in Fig.\ref{all}.



We searched for the presence of the pulsations at $\sim$34 ms in the data of 2000 and 2001 (epochs B and C), that were not analyzed by \citet{kaa00}. Following these authors, we used the top-layer data in the energy range 4.4--23.6 keV collected in Good Xenon format. The arrival times were barycentered using the position of the optical counterpart and analysed using the Rayleigh test statistics. To take into account a possible spin-up or spin-down of the pulsar since the time of the previous period measurement (P = 33.9 ms in August-September 1999), we searched for periods in a relatively large range, from 33.4 to 34.4 ms. This corresponds to a conservative assumption of \textbar$\pdot$\textbar $\sim 10^{-11}$ s s$^{-1}$ for the maximum spin-up/spin-down of the source. We performed the period search in each observation separately, without finding any significant excess in the power distributions; in all cases we found distribution peaks with power $Z^2 < 23$, corresponding to a chance detection probability $P >$  3.8 \%. We searched for excess power also in the periodograms summed over groups of observations, again with negative results. Note, however, that the frequency modulation induced by orbital motion reduces the sensitivity of this analysis. Unfortunately the orbital parameters derived by \citet{kaa00} are not sufficiently precise to correct the arrival times for the orbital motion. 
After accounting for the large contribution of the background counts, we estimate that the single observations give a sensitivity to pulsed fractions of the order of 40-60 \%, with the exact value depending on their duration and source flux.

\section{Discussion}

\begin{figure*}[t!]
\begin{center}
\resizebox{\hsize}{!}{\includegraphics[angle=-90,clip=true]{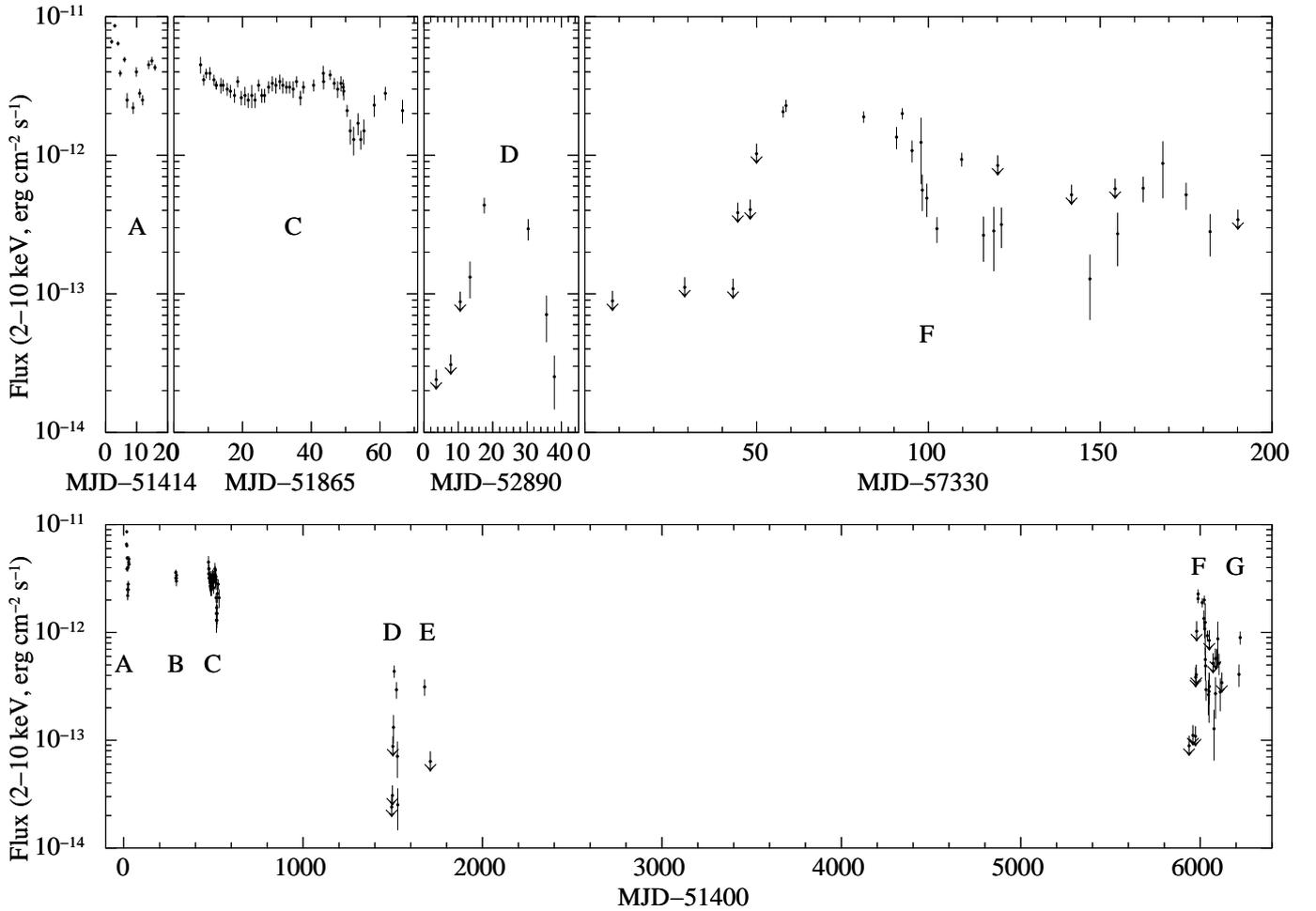}}
\end{center}
\caption{\footnotesize{Unasbsorbed 2-10 keV flux of \saxj\ measured during the observations perfomed with \RXTE\ (epochs A, B, and C), \XMM\ (epochs D and E), and \Swift\ (epochs F and G) between 1999 and 2016. \textit{Upper panels:} source light curve for the observations of epochs A, C, D, and F. \textit{Lower panel:} overall light curve since MJD = 51400 (August 10th, 1999).}}
\label{all}
\end{figure*}

In Fig.~\ref{all} (\textit{lower panel}) we report the long-term light curve of \saxj\ since MJD = 51400 (August 10th, 1999); it includes all the flux measurements obtained after the source discovery in 1997 with different telescopes. In the four \textit{upper panels} we show a zoom into the observations of epochs A, C, D, and F, respectively; for each of them we use the same scale for the time and flux axes, in order to better compare the flux variability among the different epochs. The light-curve shows that \saxj\ was clearly detected in most of the observations, although at very different flux levels. The source is very variabile, with a dynamic range of more than two orders of magnitude; moreover, this variability is also rather fast, since the source flux can increase/decrease of one order of magnitude over a timescale of a few days. However, in spite of this high variability, we never observed again an outburst as luminous as that seen with \SAX\ in 1997. Therefore, this type of event must be very rare.

In most observations the source was detected at a flux level $f_{\rm X} \sim 10^{-13} - 10^{-12}$ erg cm$^{-2}$ s$^{-1}$, corresponding to $L_{\rm X} \sim 10^{32} - 10^{33}$ erg s$^{-1}$. In the latest years similar quiescent luminosities have been observed in several  Be X-ray binaries, and various emission mechanisms have been proposed to explain them \citep[see e.g.][]{Tsygankov+17a}. In the case of \saxj, we can exclude that the observed low-luminosity originates from the companion: in fact, although the intrinsic X-ray luminosity of Be stars can be as high as $\sim 10^{32}$ erg s$^{-1}$ \citep{Naze09,Naze+11}, these stars do not show the high and fast variability observed in \saxj. For the same reason also thermal emission from the neutron star due to crustal heating in previous outbursts cannot be considered. Moreover, based on the relation between quiescent luminosity and average accretion rate, $L_{\rm q}$ = ($\dot M/10^{-11} \msun$ yr$^{-1}$)$\times 6 \times 10^{32}$ erg s$^{-1}$ \citep{Tsygankov+17a}, we would expect $L_{\rm q} \lsim 10^{31}$ erg s$^{-1}$, a value well below those observed in \saxj. Therefore, we conclude that the source emission is most likely due to matter accretion. However, the very short pulse period of 33.8 ms puts several constraints of the accretion regime: on one hand, it hampers any possibility of accretion from a cold recombined disc, regardless of the magnetic field \citep{Tsygankov+17b}; on the other hand, it makes very unlikely the subsonic accretion, even in the case of plasma radiative cooling \citep{Shakura+13}. For a typical neutron star magnetic field ($B \sim 10^{12}$) the low source luminosity can be explained only with a propeller regime, where the accreting matter is stopped by the centrifugal barrier at the magnetosphere: this is true not only for the low luminosity states, but even for the outburst observed in 1997.

\begin{figure*}
\begin{center}
\resizebox{\hsize}{!}{\includegraphics[angle=-90,clip=true]{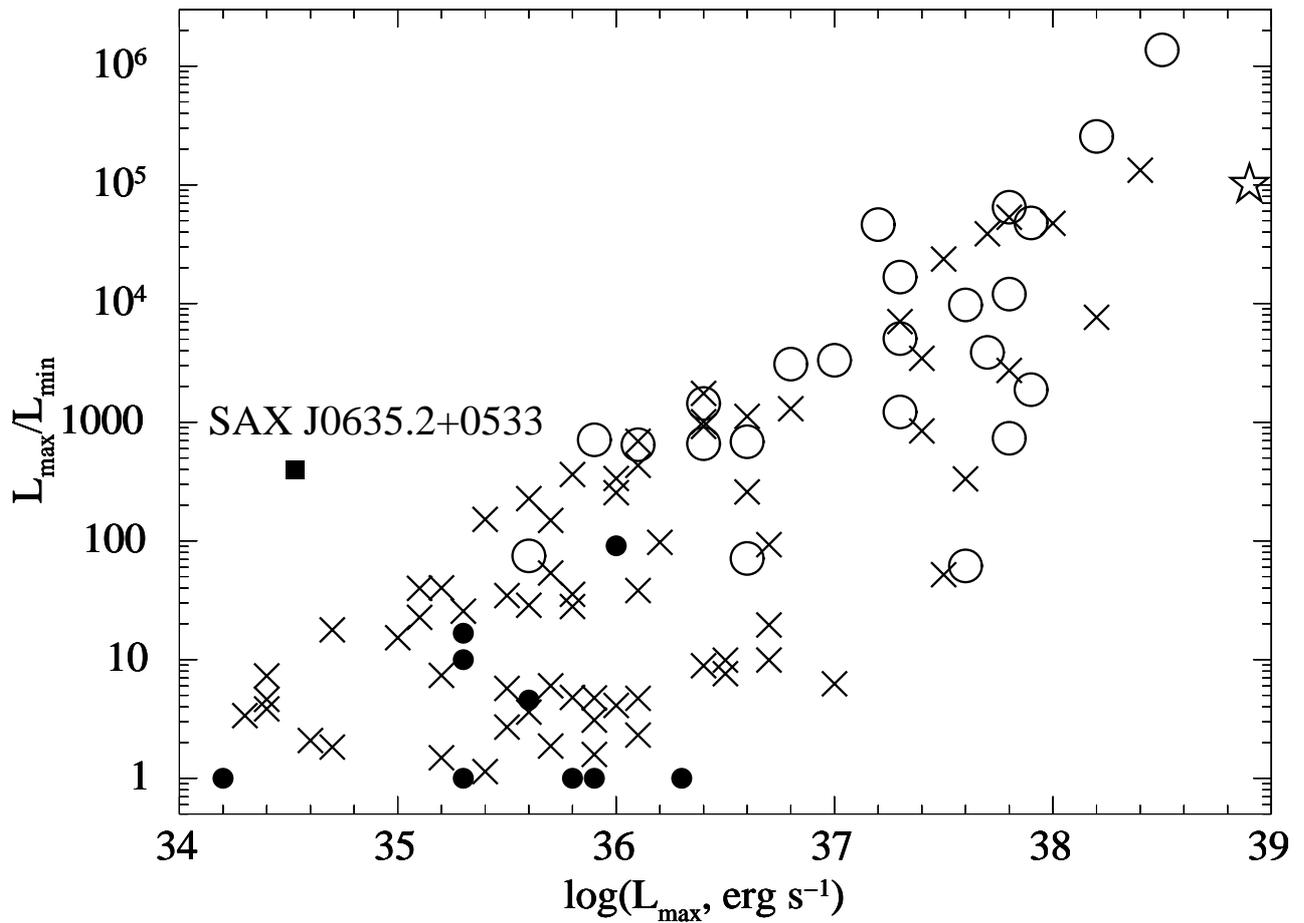}}
\end{center}
\caption{\footnotesize{X-ray luminosity ratio $L_{\rm max}/L_{\rm min}$ as a function of $L_{\rm max}$, for \saxj\ (\textit{filled square}), A0538-66 (\textit{star}), the persistent BeXRBs (\textit{filled circles}), and the transient BeXRBs in the MW (\textit{open circles}) and in the SMC (\textit{crosses}).}}
\label{bexrbs}
\end{figure*}

A large flux variability is rather unusual for a low-luminosity BeXRB such as \saxj, where the maximum flux observed in 1997 implies $L_{\rm max} \simeq 3.4\times10^{34}$ erg s$^{-1}$. From this point of view, it is interesting to compare this source with other known BeXRBs, both in the Milky Way (MW) and in the Small Magellanic Cloud (SMC). In Fig.~\ref{bexrbs} we report the $L_{\rm max}/L_{\rm min}$ ratio as a function of $L_{\rm max}$, for several known BeXRBs: \textit{open circles} represent the transient MW sources \citep{Tsygankov+17a}, while the \textit{crosses} refer to the SMC sources reported by \citet{HaberlSturm16}; in addition, \textit{filled circles} represent the class of \textit{persistent}, long spin period and low-luminosity BeXRBs \citep[see e.g.][]{LaPalombara+12}; finally, we report (as \textit{star}) also the short-period ($P$ = 69 ms) binary pulsar in the LMC A0538-66 \citep{ski82,Kretschmar+04}. The figure shows that the variability properties as a function of maximum luminosity for the MW and MC sources are similar: in both cases the source dynamic range increases with the maximum luminosity, while the low-luminosity sources are also less variable (as in the case of the persistent sources).
The lack of persistent high-luminosity sources (lower-right corner of the figure) is due to the transient nature of the BeXRBs which show a high luminosity only during the periastron passage of the NS (type I outbursts) or during a massive matter ejection from the Be star (type II outbursts).

In Fig.~\ref{bexrbs}, \saxj\ (indicated as a \textit{filled square}) is clearly an outlier: it shows the largest dynamic range ($L_{\rm max}/L_{\rm min} \simeq 400$) among the less luminous sources. Although the lack of SMC sources in the same region of the plot might be due to the difficulty of detecting low luminosity sources at such a large distance\footnote{for SMC sources, the minimum detected flux is $f_{\rm min} \lsim 10^{-14}$ erg cm$^{-2}$ s$^{-1}$, which implies $L_{\rm min} \lsim 4\times10^{33}$ erg s$^{-1}$.}, this caveat does not apply to the sources in the Milky Way.
%
This confirms that \saxj\ has rather  peculiar properties at variance with those of the normal accretion-powered BeXRBs.

\begin{acknowledgements}
We acknowledge the constant support of the \Swift\ team for the execution of the source monitoring between 2015 and 2016.
\end{acknowledgements}

\bibliographystyle{aa}
\bibliography{SAX-J0635_XRT_I2}

\onecolumn
\begin{appendix}

\section{Summary of the \Swift\ and \RXTE\ observations}
\vspace{-0.5cm}

\begin{table*}[h!]
\caption{Main parameters of the source observations perfomed with \Swift; observations of epochs F and G are separated by a horizontal line.} \label{observations_swift}
\begin{tabular}{cccccc}
 \hline
Observation	& Start Observation	& Net exposure	& CR					& Flux (0.3-10 keV)				& Luminosity$^{(a)}$ (0.3-10 keV)	\\
Number		& date - UT		& (ks)		& ($\times10^{-3}$ cts s$^{-1}$)	& ($\times10^{-13}$ erg cm$^{-2}$ s$^{-1}$)	& ($\times10^{33}$ erg s$^{-1}$)	\\ \hline
1		& 2015-11-12 - 00:44:56	& 9.3		& $<$ 1.4				& $<$ 1.0					& $<$ 0.6				\\
2		& 2015-12-03 - 00:59:16	& 9.7		& $<$ 1.8				& $<$ 1.2					& $<$ 0.8				\\
3		& 2015-12-17 - 03:05:54	& 6.0		& $<$ 1.7				& $<$ 1.2					& $<$ 0.8				\\
4		& 2015-12-18 - 12:44:49	& 1.9		& $<$ 6.1				& $<$ 4.2					& $<$ 2.8				\\
5		& 2015-12-22 - 02:44:50	& 3.8		& $<$ 6.5				& $<$ 4.4					& $<$ 2.9				\\
6		& 2015-12-23 - 23:29:48	& 0.9		& $<$ 16.4				& $<$ 11.1					& $<$ 7.3				\\
7		& 2015-12-31 - 16:23:48	& 5.2		& 33$\pm$3				& 22$\pm$2					& 15$\pm$1				\\
8		& 2016-01-01 - 13:09:41	& 3.8		& 36$\pm$3				& 25$\pm$2					& 16$\pm$2				\\
9		& 2016-01-24 - 03:39:37	& 5.2		& 30$\pm$3				& 21$\pm$2					& 13$\pm$1				\\
10		& 2016-02-02 - 17:37:57	& 1.9		& 22$\pm$4				& 15$\pm$3					& 10$\pm$2				\\
11		& 2016-02-04 - 09:27:21	& 5.0		& 32$\pm$3				& 22$\pm$2					& 14$\pm$1				\\
12		& 2016-02-07 - 05:52:29	& 2.2		& 17$\pm$3				& 12$\pm$2					& 8$\pm$1				\\
13		& 2016-02-09 - 20:37:15	& 0.4		& 20$\pm$10				& 14$\pm$7					& 9$\pm$4				\\
14		& 2016-02-10 - 05:43:32	& 1.8		& 9$\pm$3				& 6$\pm$2					& 4$\pm$1				\\
15		& 2016-02-11 - 13:38:18	& 3.1		& 8$\pm$2				& 5$\pm$1					& 3.5$\pm$0.9				\\
16		& 2016-02-14 - 11:50:20	& 7.1		& 5$\pm$1				& 3.2$\pm$0.7					& 2.1$\pm$0.4				\\
17		& 2016-02-21 - 16:07:04	& 8.1		& 15$\pm$2				& 10$\pm$1					& 6.7$\pm$0.7				\\
18		& 2016-02-28 - 01:12:03	& 2.9		& 4$\pm$1				& 3$\pm$1					& 1.9$\pm$0.7				\\
19		& 2016-03-02 - 01:00:09	& 1.5		& 5$\pm$2				& 3$\pm$1					& 2$\pm$1				\\
20		& 2016-03-03 - 04:09:49	& 0.8		& $<$ 13				& $<$ 9						& $<$ 6					\\
21		& 2016-03-04 - 05:38:50	& 4.6		& 5$\pm$2				& 3$\pm$1					& 2.3$\pm$0.7				\\
22		& 2016-03-24 - 15:32:05	& 3.4		& $<$ 8					& $<$ 6						& $<$ 4					\\
23		& 2016-03-30 - 00:45:50	& 4.2		& 2$\pm$1				& 1.4$\pm$0.7					& 0.9$\pm$0.4				\\
24		& 2016-04-06 - 08:03:15	& 1.7		& $<$ 9					& $<$ 6						& $<$ 4					\\
25		& 2016-04-07 - 01:38:57	& 2.1		& 4$\pm$2				& 3$\pm$1					& 1.9$\pm$0.8				\\
26		& 2016-04-14 - 12:15:17	& 3.7		& 9$\pm$2				& 6$\pm$1					& 4.1$\pm$0.8				\\
27		& 2016-04-20 - 05:26:58	& 0.5		& 14$\pm$6				& 9$\pm$4					& 6$\pm$3				\\
28		& 2016-04-21 - 15:15:29	& 0.1		& -					& -						& -					\\
29		& 2016-04-27 - 00:13:19	& 3.6		& 8$\pm$2				& 6$\pm$1					& 3.7$\pm$0.8				\\
30		& 2016-05-04 - 00:03:16	& 3.2		& 4$\pm$1				& 3$\pm$1					& 2.0$\pm$0.7				\\
31		& 2016-05-12 - 02:28:06	& 4.4		& $<$5					& $<$4						& $<$2					\\ \hline
32		& 2016-08-16 - 06:16:08	& 3.9		& 7$\pm$2				& 4$\pm$1					& 2.9$\pm$0.7				\\
33		& 2016-08-23 - 04:06:43	& 5.0		& 14$\pm$2				& 10$\pm$1					& 6.4$\pm$0.9				\\ \hline
\end{tabular}
\\
$^{(a)}$ Corrected for the absorption, and assuming a distance of 5 kpc.
\vspace{-5cm}
\end{table*}
\vspace{-5cm}

\begin{table*}
\vspace{-5cm}
\caption{Main parameters of the source observations perfomed with \RXTE; observations of epochs A, B, and C are separated by a horizontal line.}
\label{observations_rxte}
\begin{tabular}{ccccccc}
\hline
Observation	& Observation		& Start Observation	& Net exposure	& Flux (2-10 keV)				& Luminosity$^{(a)}$ (2-10 keV)		\\
number		& ID			& date - UT		& (s)		& ($\times10^{-12}$ erg cm$^{-2}$ s$^{-1}$)	& ($\times10^{33}$ erg s$^{-1}$)	\\ \hline
01		& 40131-01-01-00	& 1999-08-25 22:10:06.8	& 2896		& 6.6$\pm$0.2					& 22.0	$\pm$0.7			\\
02		& 40131-01-02-00	& 1999-08-26 22:06:28.2	& 2880		& 8.6$^{+0.2}_{-0.3}$				& 28.7	$^{+0.7}_{-1.0}$		\\
03		& 40131-01-03-00	& 1999-08-27 23:43:50.6	& 2640		& 6.4$^{+0.2}_{-0.3}$				& 21.3	$^{+0.7}_{-1.0}$		\\
04		& 40131-01-04-00	& 1999-08-28 17:15:50.1	& 2800		& 3.9$^{+0.2}_{-0.3}$				& 13.0	$^{+0.7}_{-1.0}$		\\
05		& 40131-01-05-00	& 1999-08-30 01:12:46.7	& 2368		& 4.9$^{+0.2}_{-0.3}$				& 16.3	$^{+0.7}_{-1.0}$		\\
06		& 40131-01-06-00	& 1999-08-30 21:57:38.6	& 2944		& 2.5$^{+0.3}_{-0.2}$				& 8.3	$^{+1.0}_{-0.7}$		\\
07		& 40131-01-07-00	& 1999-09-01 20:35:06.1	& 2592		& 2.2$^{+0.2}_{-0.3}$				& 7.3	$^{+0.7}_{-1.0}$		\\
08		& 40131-01-08-00	& 1999-09-02 21:48:36.9	& 2928		& 4.0$^{+0.3}_{-0.2}$				& 13.3	$^{+1.0}_{-0.7}$		\\
09		& 40131-01-09-00	& 1999-09-03 23:32:07.3	& 2816		& 2.8$\pm$0.2					& 9.3	$\pm$0.7			\\
10		& 40131-01-10-00	& 1999-09-04 21:49:58.9	& 3120		& 2.5$\pm$0.2					& 8.3	$\pm$0.7			\\
11		& 40131-01-11-00	& 1999-09-06 21:51:07.2	& 3120		& 4.5$^{+0.3}_{-0.2}$				& 15.0	$^{+1.0}_{-0.7}$		\\
12		& 40131-01-12-00	& 1999-09-07 21:42:20.2	& 3184		& 4.8$\pm$0.3					& 16.0	$\pm$1.0			\\
13		& 40131-01-13-00	& 1999-09-08 21:47:07.0	& 3168		& 4.3$\pm$0.2					& 14.3	$\pm$0.7			\\ \hline
\end{tabular}
\\
$^{(a)}$ Corrected for the absorption, and assuming a distance of 5 kpc.
\end{table*}

\addtocounter{table}{-1}
\begin{table*}
\caption{Continued}
\label{observations}
\begin{tabular}{ccccccc}
\hline
Observation	& Observation		& Start Observation	& Net exposure	& Flux (2-10 keV)				& Luminosity$^{(a)}$ (2-10 keV)		\\
number		& ID			& date - UT		& (ks)		& ($\times10^{-12}$ erg cm$^{-2}$ s$^{-1}$)	& ($\times10^{33}$ erg s$^{-1}$)	\\ \hline
14		& 50085-01-01-00	& 2000-05-26 04:18:08.9	& 6064		& 3.6$\pm$0.2					& 12.0	$\pm$0.7			\\
15		& 50085-01-02-00	& 2000-05-27 02:42:24.2	& 6480		& 3.2$\pm$0.2					& 10.7	$\pm$0.7			\\
16		& 50085-01-03-00	& 2000-05-28 02:58:28.4	& 5704		& 3.2$\pm$0.2					& 10.7	$\pm$0.7			\\
17		& 50085-01-04-00	& 2000-05-29 04:09:25.3	& 3248		& 3.0$^{+0.3}_{-0.2}$				& 10.0	$^{+1.0}_{-0.7}$		\\
18		& 50085-01-05-00	& 2000-05-30 05:36:15.3	& 2832		& 3.4$^{+0.3}_{-0.2}$				& 11.3	$^{+1.0}_{-0.7}$		\\ \hline
19		& 50085-01-06-00	& 2000-11-24 18:36:26.8	& 1960		& 4.5$^{+0.6}_{-0.3}$				& 15.0	$^{+2.0}_{-1.0}$		\\
20		& 50085-01-07-00	& 2000-11-25 16:47:27.2	& 2488		& 3.5$^{+0.3}_{-0.2}$				& 11.7	$^{+1.0}_{-0.7}$		\\
21		& 50085-01-08-00	& 2000-11-26 10:18:51.3	& 3104		& 3.9$^{+0.3}_{-0.4}$				& 13.0	$^{+1.0}_{-1.3}$		\\
22		& 50085-01-09-00	& 2000-11-27 11:48:54.7	& 2352		& 3.9$^{+0.4}_{-0.3}$				& 13.0	$^{+1.3}_{-1.0}$		\\
23		& 50085-01-10-00	& 2000-11-28 16:28:57.8	& 2016		& 3.5$^{+0.3}_{-0.4}$				& 11.7	$^{+1.0}_{-1.3}$		\\
24		& 50085-01-11-00	& 2000-11-29 08:24:00.9	& 2432		& 3.2$^{+0.2}_{-0.3}$				& 10.7	$^{+0.7}_{-1.0}$		\\
25		& 50085-01-12-00	& 2000-11-30 16:18:04.6	& 2000		& 3.2$^{+0.4}_{-0.3}$				& 10.7	$^{+1.3}_{-1.0}$		\\
26		& 50085-01-13-00	& 2000-12-01 08:10:08.8	& 3264		& 3.2$\pm$0.3					& 10.7	$\pm$1.0			\\
27		& 50085-01-14-00	& 2000-12-02 12:51:11.5	& 3024		& 3.0$\pm$0.3					& 10.0	$\pm$1.0			\\
28		& 50085-01-15-00	& 2000-12-03 11:08:20.8	& 2664		& 2.9$\pm$0.3					& 9.7	$\pm$1.0			\\
29		& 50085-01-16-00	& 2000-12-04 15:50:21.4	& 2032		& 2.7$^{+0.3}_{-0.4}$				& 9.0	$^{+1.0}_{-1.3}$		\\
30		& 50085-01-17-00	& 2000-12-05 14:04:06.8	& 2448		& 3.4$^{+0.3}_{-0.4}$				& 11.3	$^{+1.0}_{-1.3}$		\\
31		& 50085-01-18-00	& 2000-12-06 15:36:32.0	& 2544		& 2.6$\pm$0.3					& 8.7	$\pm$1.0			\\
32		& 50085-01-19-00	& 2000-12-07 17:07:44.5	& 2448		& 2.7$^{+0.4}_{-0.3}$				& 9.0	$^{+1.3}_{-1.0}$		\\
33		& 50085-01-20-00	& 2000-12-08 15:22:41.7	& 2800		& 2.5$\pm$0.3					& 8.3	$\pm$1.0			\\
34		& 50085-01-21-00	& 2000-12-09 16:53:56.8	& 2528		& 2.7$^{+0.5}_{-0.3}$				& 9.0	$^{+1.7}_{-1.0}$		\\
35		& 50085-01-22-00	& 2000-12-10 15:07:56.1	& 2880		& 2.5$\pm$0.3					& 8.3	$\pm$1.0			\\
36		& 50085-01-23-00	& 2000-12-11 16:39:04.3	& 2608		& 3.2$\pm$0.3					& 10.7	$\pm$1.0			\\
37		& 50085-01-24-00	& 2000-12-12 14:54:09.2	& 2944		& 2.7$\pm$0.3					& 9.0	$\pm$1.0			\\
38		& 50085-01-25-00	& 2000-12-13 10:01:16.3	& 2704		& 2.7$^{+0.3}_{-0.4}$				& 9.0	$^{+1.0}_{-1.3}$		\\
39		& 50085-01-26-00	& 2000-12-14 14:44:05.3	& 3040		& 3.1$\pm$0.3					& 10.3	$\pm$1.0			\\
40		& 50085-01-27-00	& 2000-12-15 16:10:35.3	& 2784		& 3.3$^{+0.4}_{-0.3}$				& 11.0	$^{+1.3}_{-1.0}$		\\
41		& 50085-01-28-00	& 2000-12-16 17:41:46.2	& 2528		& 3.2$^{+0.4}_{-0.3}$				& 10.7	$^{+1.3}_{-1.0}$		\\
42		& 50085-01-29-00	& 2000-12-17 20:46:32.2	& 1768		& 3.4$\pm$0.4					& 11.3	$\pm$1.3			\\
43		& 50085-01-30-00	& 2000-12-18 17:27:43.8	& 2608		& 3.2$\pm$0.4					& 10.7	$\pm$1.3			\\
44		& 50085-01-31-00	& 2000-12-19 18:53:40.1	& 2336		& 3.1$\pm$0.3					& 10.3	$\pm$1.0			\\
45		& 50085-01-32-00	& 2000-12-20 17:07:39.4	& 2672		& 3.1$\pm$0.3					& 10.3	$\pm$1.0			\\
46		& 50085-01-33-00	& 2000-12-21 18:40:44.3	& 2416		& 3.0$^{+0.4}_{-0.3}$				& 10.0	$^{+1.3}_{-1.0}$		\\
47		& 50085-01-34-00	& 2000-12-22 16:55:42.2	& 2736		& 3.4$^{+0.3}_{-0.4}$				& 11.3	$^{+1.0}_{-1.3}$		\\
48		& 50085-01-35-00	& 2000-12-23 20:02:39.6	& 2664		& 2.6$\pm$0.3					& 8.7	$\pm$1.0			\\
49		& 50085-01-36-00	& 2000-12-24 18:20:17.4	& 3248		& 3.1$\pm$0.3					& 10.3	$\pm$1.0			\\
50		& 50085-01-37-00	& 2000-12-27 16:19:45.7	& 2896		& 3.2$\pm$0.3					& 10.7	$\pm$1.0			\\
51		& 50085-01-38-00	& 2000-12-30 12:47:45.9	& 1528		& 3.9$^{+0.5}_{-0.4}$				& 13.0	$^{+1.7}_{-1.3}$		\\
52		& 50085-01-38-01	& 2000-12-30 14:31:56.9	& 1608		& 3.4$\pm$0.4					& 11.3	$\pm$1.3			\\
53		& 50085-01-39-00	& 2001-01-01 12:34:12.9	& 3296		& 3.8$\pm$0.3					& 12.7	$\pm$1.0			\\
54		& 50085-01-40-00	& 2001-01-02 16:21:06.0	& 4512		& 3.3$\pm$0.3					& 11.0	$\pm$1.0			\\
55		& 50085-01-41-00	& 2001-01-03 16:17:06.7	& 2728		& 3.0$^{+0.4}_{-0.3}$				& 10.0	$^{+1.3}_{-1.0}$		\\
56		& 50085-01-42-00	& 2001-01-04 15:32:15.4	& 2736		& 3.3$^{+0.4}_{-0.3}$				& 11.0	$^{+1.3}_{-1.0}$		\\
57		& 50085-01-40-01	& 2001-01-05 07:24:44.6	& 1872		& 3.1$\pm$0.4					& 10.3	$\pm$1.3			\\
58		& 50085-01-43-00	& 2001-01-05 10:48:27.6	& 2112		& 2.9$\pm$0.4					& 9.7	$\pm$1.3			\\
59		& 50085-01-44-00	& 2001-01-06 10:24:52.4	& 2912		& 2.1$^{+0.2}_{-0.3}$				& 7.0	$^{+0.7}_{-1.0}$		\\
60		& 50085-01-45-00	& 2001-01-07 08:38:26.6	& 2448		& 1.5$^{+0.3}_{-0.2}$				& 5.0	$^{+1.0}_{-0.7}$		\\
61		& 50085-01-46-00	& 2001-01-08 08:31:10.3	& 2416		& 1.3$^{+0.3}_{-0.2}$				& 4.3	$^{+1.0}_{-0.7}$		\\
62		& 50085-01-47-00	& 2001-01-09 14:56:14.5	& 2848		& 1.7$^{+0.3}_{-0.2}$				& 5.7	$^{+1.0}_{-0.7}$		\\
63		& 50085-01-48-00	& 2001-01-10 09:57:12.7	& 2816		& 1.3$\pm$0.2					& 4.3	$\pm$0.7			\\
64		& 50085-01-49-00	& 2001-01-11 08:15:47.5	& 2368		& 1.5$^{+0.3}_{-0.2}$				& 5.0	$^{+1.0}_{-0.7}$		\\
65		& 50085-01-50-00	& 2001-01-14 07:49:49.7	& 2336		& 2.3$^{+0.4}_{-0.3}$				& 7.7	$^{+1.3}_{-1.0}$		\\
66		& 50085-01-51-00	& 2001-01-17 13:57:52.7	& 2000		& 2.8$^{+0.3}_{-0.5}$				& 9.3	$^{+1.0}_{-1.7}$		\\
67		& 50085-01-52-00	& 2001-01-22 13:24:36.9	& 2000		& 2.1$^{+0.4}_{-0.3}$				& 7.0	$^{+1.3}_{-1.0}$		\\ \hline
\end{tabular}
\end{table*}

\end{appendix}

\end{document}